\documentclass[review]{elsarticle}

\usepackage{amsmath,amssymb,bm}
\usepackage{physics}
\usepackage{color}
\usepackage{comment}

\usepackage{lineno,hyperref}
\usepackage{booktabs}
\usepackage{multirow}

\newcommand{\LA}{\mathop{\mathcal{L}}\nolimits}

\newcommand{\R}{\mathop{\mathbb{R}}\nolimits}
\newcommand{\CA}{\mathop{\mathbb{C}}\nolimits}

\newcommand{\I}{\mathop{\mathbb{I}}\nolimits}

\newcommand{\ct}{\mathrm{\ast}} 
\newcommand{\ov}[1]{\overline{#1}}

\newcommand{\bracket}[2]{\langle #1,#2 \rangle}

\newtheorem{Theorem}{Theorem}
\newtheorem{Proposition}{Proposition}

\newtheorem{Lemma}{Lemma}

\begin{document}

\begin{frontmatter}

\title{Constraints for the spectra of generators of  quantum dynamical semigroups}

\author[1]{Dariusz Chru\'sci\'nski}
\author[2]{Ryohei Fujii}
\author[2]{Gen Kimura \corref{C}}
\ead{gen@shibaura-it.ac.jp}
\author[3]{Hiromichi Ohno}

\address[1]{Institute of Physics, Faculty of Physics, Astronomy and Informatics  Nicolaus Copernicus University, Grudzi\c{a}dzka 5/7, 87--100 Toru\'n, Poland}
\address[2]{College of Systems Engineering and Science,
Shibaura Institute of Technology, Saitama 330-8570, Japan}
\address[3]{Department of Mathematics, Faculty of Engineering, Shinshu University,
4-17-1 Wakasato, Nagano 380-8553, Japan.} 
\cortext[C]{Corresponding author}

\begin{abstract}
Motivated by a spectral analysis of the generator of completely positive trace-preserving semigroup, we analyze a real functional
$$
A,B \in M_n(\CA) \to r(A,B) = \frac{1}{2}\Bigl(\bracket{[B,A]}{BA} + \bracket{[B,A^\ct]}{BA^\ct}\Bigr) \in \R
$$ 
where $\bracket{A}{B} := \tr (A^\ct B)$ is the Hilbert-Schmidt inner product, and $[A,B]:= AB - BA$ is the commutator. In particular we discuss the upper and lower bounds of the form $c_- \|A\|^2 \|B\|^2 \le r(A,B) \le c_+ \|A\|^2 \|B\|^2$ where $\|A\|$ is the Frobenius norm. We prove that the optimal upper and lower bounds are given by $c_\pm = \frac{1 \pm \sqrt{2}}{2}$.  If $A$ is restricted to be traceless, the bounds are further improved to be $c_\pm = \frac{1 \pm \sqrt{2(1-\frac{1}{n})}}{2}$. 
Interestingly, these upper bounds, especially the latter one, provide new constraints on relaxation rates for the quantum dynamical semigroup tighter than previously known constraints in the literature. 
A relation with B\"{o}ttcher-Wenzel inequality is also discussed.    
\end{abstract}


\end{frontmatter}

\section{Introduction}

Motivated by a problem in the field of open quantum systems, we introduce the following real-valued functional: 
\begin{equation}\label{eq:rf}
r(A,B) = \frac{1}{2}\Bigl(\bracket{[B,A]}{BA} + \bracket{[B,A^\ct]}{BA^\ct}\Bigr)
\end{equation}
for any complex square matrices $A,B \in M_n(\CA)$. Here, $A^\ct$ is the Hermitian conjugation of $A$, $\bracket{A}{B} := \tr (A^\ct B)$ is the Hilbert-Schmidt inner product\footnote{Note that, we are following the convention in quantum physics that the inner product is linear in the second argument (and anti-linear in the first argument). }, and $[A,B]:= AB - BA$ denotes the commutator. 
One of the goals of this paper is to investigate the bounds of the form 
\begin{equation}\label{eq:bd}
c_-(n) \|A\|^2 \|B\|^2  \le r(A,B) \le c_+(n) \|A\|^2 \|B\|^2 
\end{equation}
where $c_\pm (n)$ are real constants (possibly dependent on the matrix size $n$) and the matrix norm is the Frobenius norm $\|A\| := \sqrt{\tr (A^\ct A)}$. 
We will see (in Sec.~\ref{sec:BW}) that these bounds problem is related to B\"{o}ttcher-Wenzel inequality \cite{ref:BW1} on the norm of commutator. 

The origin of the function \eqref{eq:rf} lies in the field of open quantum system \cite{ref:Breuer,ref:Dav,ref:Alicki,ref:GZ,ref:GKreview}: 
For a Markovian open quantum dynamics, a time evolution of quantum state is described by quantum dynamical semigroup, i.e., a semigroup of completely positive trace-preserving maps \cite{ref:Alicki}. Due to the seminal results \cite{ref:GKS,ref:L}, any such generator has the following Gorini-Kossakowski-Lindblad-Sudarshan (GKLS) form
\begin{equation}\label{DIAG}
\mathcal{L}(\rho)= - i [H,\rho] + \sum_{k} \left(L_k\rho L_k^\ct -\frac 12\{L_k^\ct L_k,\rho\}\right),
\end{equation}
where $H = H^\ast$ is interpreted as an effective Hamiltonian, and $L_k$'s are the so called jump (or noise) operators. 
One of the important physical quantities is a relaxation rate which determines the time scale of the exponential decaying \cite{ref:Alicki}. 
For $n$-level quantum system, there are generally $n^2-1$ relaxation rates, which are defined by    
\begin{equation}\label{eq:DefRC}
\Gamma_\alpha := - \Re \lambda_\alpha, \ \ \ \ (\alpha = 1,\ldots, n^2-1),  
\end{equation}
where $\lambda_\alpha$ are eigenvalues of generator $\mathcal{L}$ except for $\lambda_{n^2} = 0$. (Note that there is always one zero eigenvalue due to the trace preserving property of $\LA$.) 
By using \eqref{eq:rf} and GKLS form \eqref{DIAG}, we observe (in Sec.~\ref{sec:Phys}) that any relaxation rate is expressed as   
\begin{equation}\label{eq:RRRF}
\Gamma_\alpha = \sum_k r(u_\alpha,L_k), 
\end{equation}
where $u_\alpha$ is the unit eigenmatrix of $\LA$ belonging to $\lambda_\alpha$, i.e., $\LA(u_\alpha) = \lambda_\alpha u_\alpha \ (\|u_\alpha\| = 1)$. 
With this in mind, we shall call function \eqref{eq:rf} a {\it $r$-function}, where ``{\it r}'' stands for a {\it relaxation}. 
Moreover, it is easy to see that the upper bound in  \eqref{eq:bd} gives the following constraints for relaxation rates (see Sec.~\ref{sec:Phys} for details): 
\begin{equation}\label{eq:ConsRelCN}
\Gamma_\alpha \le \frac{c_+(n)}{n} \sum_{\beta = 1}^{n^2-1} \Gamma_\beta \ \ \ \ (\forall \alpha = 1,\ldots,n^2-1).
\end{equation}
This is universally satisfied for any quantum dynamical semigroup reflecting the condition of complete positivity. 
Therefore, we have a significant motivation to investigate function \eqref{eq:rf}, especially its upper bound, in open quantum physics. 

In this paper, we show that the best constants for \eqref{eq:bd} (i.e., the minimum of $c_+(n)$ and the maximum of $c_-(n)$) are $c_\pm(n) = \frac{1\pm \sqrt{2}}{2}$ independent of the matrix size $n$. 
Namely, we have 
\begin{Theorem}\label{Thm:Main1New}
For any complex matrices $A,B \in M_n(\CA)$, 
\begin{equation}\label{eq:bdMain1}
\frac{1-\sqrt{2}}{2} \|A\|^2 \|B\|^2 \le r(A,B) \le \frac{1+\sqrt{2}}{2} \|A\|^2 \|B\|^2.   
\end{equation}
Both lower and upper bounds are sharp, i.e., there are matrices $A$ and $B$ that can attain the bounds. 
\end{Theorem}
Furthermore, if we restrict the matrix $A$ to be traceless, then the best constants are further tightened to be $c_\pm(n) = \frac{1 \pm \sqrt{2(1- \frac{1}{n})}}{2}$: 
\begin{Theorem}\label{Thm:Main2New}
For any complex matrices $A,B \in M_n(\CA)$ with $\tr A = 0$, 
\begin{equation}\label{eq:bdMain2}
 \frac{1 - \sqrt{2(1- \frac{1}{n})}}{2}\|A\|^2 \|B\|^2 \le r(A,B) \le \frac{1 + \sqrt{2(1- \frac{1}{n})}}{2}\|A\|^2 \|B\|^2. 
\end{equation}
Both lower and upper bounds are sharp. 
In particular, for $n=2$, the inequalities read  
\begin{equation}\label{eq:bdMain2n=2}
 0 \le r(A,B) \le \|A\|^2 \|B\|^2. 
\end{equation}
\end{Theorem}
Note that the best constants $\frac{1 \pm \sqrt{2(1- \frac{1}{n})}}{2}$ in \eqref{eq:bdMain2} with the traceless restriction are strictly tighter than the best constants $\frac{1\pm \sqrt{2}}{2}$ in \eqref{eq:bdMain1} and converges to the general bounds as $n \to \infty$. 
We emphasize that both upper bounds in \eqref{eq:bdMain1} and \eqref{eq:bdMain2} give tighter constraints than any of the previously known constraints \cite{ref:KAW,ref:CKKS} 
 (Theorem \ref{Thm:MAIN}). 

The rest of the paper is organized as follows. In Sec.~\ref{sec:PropR}, we investigate general properties of $r$-function. 
In Sec.~\ref{sec:BW}, we discuss a relation between our problem and B\"{o}ttcher-Wenzel inequality on the norm of commutator. 
In Sec.~\ref{sec:br}, we prove main results on the sharp bounds of $r$-function. In Sec.~\ref{sec:Phys} an application to quantum dynamical semigroup is explained. 
Sec.~\ref{sec:CD} is devoted to conclusion and discussion. 



\section{Properties of $r$-function}\label{sec:PropR}
In this section, we investigate some of general properties of $r$-function \eqref{eq:rf}. 
We first observe that there are several equivalent expressions:
\begin{eqnarray}
 r(A,B) &=& \frac{1}{2}\tr (A^\ct A B^\ct B +  A A^\ct B^\ct B - A^\ct B A B^\ct - B A^\ct B^\ct A), \label{eq:r-expand} \\ 
 &= &\frac{1}{2}\tr (\{A, A^\ct\}  B^\ct B)  - \Re \tr (A^\ct B A B^\ct),  \label{eq:g2} \\
 &=& \frac{1}{2}\Bigl((\| [A,B]\|^2 + \tr A^\ct A [B^\ct,B]\Bigr),  \label{eq:IDg3} \\
&=& \frac{1}{2}\Bigl((\|[A^\ct,B^\ct]\|^2 + \tr (A^\ct A [B^\ct,B]) \Bigr), \label{eq:IDg4}  \\
 &=& \frac{1}{2}\Bigl((\|[A,B^\ct]\|^2 + \tr (A A^\ct [B^\ct,B])\Bigr), \\
&=& \frac{1}{2}\Bigl((\|[A^\ct,B]\|^2 + \tr (A A^\ct [B^\ct,B])\Bigr), \\
 &=& \frac{1}{4}\Bigl((\|[A,B]\|^2 + \|[A^\ct,B]\|^2 + \tr(\{A,A^\ct\} [B^\ct,B])\Bigr),  \label{eq:IDg}
\end{eqnarray}
where $\{A,B\} := AB + BA$ denotes the anti-commutator. Indeed, it is straightforward to see that expansions of \eqref{eq:rf} and \eqref{eq:g2} - \eqref{eq:IDg} all reduce to \eqref{eq:r-expand} by noting the cyclic property of trace. 
The fact that $r$-function is a real valued is easily seen from these expressions. 
Note that $r$-function is asymmetric between $A$ and $B$ in general; for instance, with $A= \left(
\begin{array}{cc}
 0 & 1 \\
 1 & 1 \\
\end{array}
\right)$ and $B = \left(
\begin{array}{cc}
 0 & 0 \\
 1 & 0 \\
\end{array}
\right)$, $r(A,B) = 1$ and $r(B,A) = \frac{3}{2}$.
It is not bilinear but satisfies
\begin{equation}\label{eq:QD}
r(\alpha A, \beta B) = |\alpha|^2 |\beta|^2 r(A,B) \quad (\alpha, \beta \in \CA).
\end{equation}
Note also the unitary invariance:
\begin{equation}\label{eq:UInv}
r(UAU^\ct, U B U^\ct) = r(A,B)
\end{equation}
for any unitary matrix $U$.
For Cartesian decomposition for $A = A_R + i A_I$, where $A_R = \frac{A+A^\ast}{2}$ and $A_I = \frac{A-A^\ast}{2i}$ are self-adjoint real and imaginary parts of $A$, a direct computation shows
\begin{equation}\label{eq:CDforA}
r(A,B) = r(A_R,B) + r(A_I,B).
\end{equation}
Late on, we will see this property plays an important role in showing upper and lower bounds for general matrices, not necessarily for self-adjoint matrices.  

If $A = B$, $r$-function is shown to be (by using \eqref{eq:IDg3} or \eqref{eq:IDg4})
\begin{equation}\label{eq:raa}
r(A,A) = \frac{1}{2} \tr (A^\ct A [A^\ct,A]). 
\end{equation}
For the sake of practice, we shall start from the lower and upper bounds for this case.  
Since the eigenvalues of $A^\ct A$ and $AA^\ct$ are the same, we have $\|A^\ct A\| = \|AA^\ct\|$. 
By using Schwarz inequality, 
\begin{align*}
 r(A,A) &= \frac{1}{2} \tr(A^\ct AA^\ct A - A^\ct AAA^\ct)
= \frac{1}{2}\| A^\ct A\|^2 - \frac{1}{2}\bracket{A^\ct A}{AA^\ct} \\
&\ge \frac{1}{2} \| A^\ct A\|^2 - \frac{1}{2} \|A^\ct A\| \|AA^\ct \|
=0. 
\end{align*}
This lower bound is sharp since the equality is always attainable by any normal matrix $A$. 
Using the similar idea, one obtains the following upper bound:  
$$
r(A,A) \le  \frac{1}{2} \|A^\ct A\|^2 + \frac{1}{2} |\bracket{A^\ct A}{AA^\ct} | \le \frac{1}{2} \|A^\ct A\|^2  + \frac{1}{2}\|A^\ct A\| \|AA^\ct \| \le \|A\|^4.
$$
Here, we have used the triangle inequality, Schwarz inequality, the norm inequality $\| AB\| \le \| A\| \|B \|$ (which holds for any unitary invariant norm \cite{ref:Bhatia}), and $\|A^\ct \| = \|A\|$. 
However, this is not the tight bound and can be further sharpened as follows. 
In what follows, we use the Dirac notation \cite{ref:Dirac} for vector $\ket{\psi} \in \CA^n$. In particular, let $\ketbra{\psi}{\phi}$ denote the linear operator on $\CA^n$ defined by $\ketbra{\psi}{\phi} \ket{\xi} := \langle \phi|\xi \rangle \ket{\psi}$ where $\langle \phi|\xi \rangle$ is the inner product between $\ket{\phi}$ and $\ket{\xi}$. 
With this notation, the singular value decomposition of $A$ with its singular values $a_i \ge 0$ can be written as $A = \sum_{i=1}^n a_i \ketbra{b_i}{\beta_i}$ with orthonormal bases $\{\ket{b_i}\}_{i=1}^n,\{\ket{\beta_i}\}_{i=1}^n$ of $\CA^n$. Then, a direct computation of \eqref{eq:raa} shows 
\begin{align}
	r(A,A) = \frac{1}{2}\Bigl(\sum_{i=1}^{n} a^4_i - \sum_{i,j=1}^n a^2_i a^2_j |\langle b_i | \beta_j\rangle|^2\Bigr).  
\end{align}
Since $\|A\|^4 = (\sum_i a_i^2)^2 = \sum_i a_i^4 + \sum_{i \neq j} a_i^2 a_j^2$, we have 
$$
\frac{1}{2} \|A\|^4 - r(A,A) = \frac{1}{2}\Bigl( \sum_{i \neq j} a_i^2 a_j^2 + \sum_{i,j=1}^n a^2_i a^2_j |\langle b_i | \beta_j\rangle|^2 \Bigr) \ge 0. 
$$ 
The equality is achieved by a rank $1$ operator $A = a \ketbra{b}{\beta}$ with orthogonal $\ket{b}$ and $\ket{\beta}$. 
Hence, we have obtained: 
\begin{Proposition}
For any $A \in M_n(\CA)$, 
\begin{equation}\label{eq:raaSPB}
0\le r(A,A) \le \frac{1}{2} \|A\|^4.
\end{equation}
Both lower and upper bounds are sharp. 
\end{Proposition}

\section{$r$-function vs. B{\"o}ttcher--Wenzel inequality}\label{sec:BW}

Interestingly, our problem is closely related to B{\"o}ttcher--Wenzel inequality:
\begin{equation}\label{eq:BW}
\| [A,B] \|^2 \le 2 \| A\|^2 \| B \|^2, 
\end{equation}
where the bound is sharp. This inequality was firstly conjectured by B{\"o}ttcher and Wenzel in \cite{ref:BW0} giving a proof for real $2 \times 2$ matrices and also normal matrices. Then, it was proved by Vong and Jin \cite{ref:BW2}, by Lu \cite{ref: Lu}, and subsequently by B{\"o}ttcher and Wenzel themselves for general complex matrices \cite{ref:BW1}. A simple and conceptually sound proof was  given by Audenaert using a variance bound \cite{ref:BW_Var}. For the equality condition, see \cite{ref:BW1} and \cite{ref:Comm_maxFrob}. 

In regard to our problem, if we restrict the matrix $B$ to be normal, $r$-function reduces to be the norm of commutator: 
\begin{equation}\label{eq:rfBnorm}
r(A,B) =\frac{1}{2} \| [A,B] \|^2,
\end{equation}
which is easily observed from the expression \eqref{eq:IDg3}.
Since the equality in \eqref{eq:BW} can be attained by a normal matrix $B$ (see Proposition 4.6 in \cite{ref:BW1}), the best bound in \eqref{eq:bd} with the restriction of $B$ being normal is $1$.
Therefore, one observes the following restricted inequality:
\begin{equation}\label{eq:NBcase}
r(A,B) \le \|A\|^2 \|B\|^2
\end{equation}
which is satisfied for any complex matrix $A$ and for any normal matrix $B$. 
Notice that the situation is completely asymmetric between $A$ and $B$. Indeed, in the next section, we will see (Proposition \ref{Prop:1}) that the restriction for $A$ being normal (and even self-adjoint) does not change the general sharp bound \eqref{eq:bdMain1}.

Furthermore, B{\"o}ttcher--Wenzel inequality \eqref{eq:BW} gives a non-trivial upper bound for our problem \eqref{eq:bd}:
\begin{Proposition}\label{prop:BW} For any matrices $A,B \in M_n(\CA)$,
$$
r(A,B) \le \sqrt{2} \|A \|^2 \|B\|^2.
$$
\end{Proposition}
[Proof] : Applying the triangle inequality, Schwarz inequality, the norm inequality $\| AB\| \le \| A\| \|B \|$, $\|A^\ast\| = \|A\|$, and finally commutator inequality \eqref{eq:BW} to $r$-function in the form \eqref{eq:rf}, we have
\begin{eqnarray*}
r(A,B) &\le& \frac{1}{2}(|\bracket{[B,A]}{BA}| + |\bracket{[B,A^\ct]}{BA^\ct}| ) \\
&\le& \frac{1}{2}(\|[B,A]\| \|BA\| + \|  [B,A^\ct] \| \| BA ^\ct \|) \le \sqrt{2} \|A\|^2 \|B\|^2.
\end{eqnarray*}
\hfill $\blacksquare $

Notice that if we apply this result to \eqref{eq:ConsRelCN}, we recover the following universal constraints for relaxation rates: 
\begin{equation}\label{eq:ConsRelCNKAW}
\Gamma_\alpha \le \frac{\sqrt{2}}{n} \sum_{\beta = 1}^{n^2-1} \Gamma_\beta \ \ \ \ (\forall \alpha = 1,\ldots,n^2-1).  
\end{equation}
This bound was observed in \cite{ref:KAW} by one of the authors (G.K.) essentially using the same reasoning here. 
However, as we will see in the next section, the upper bounds are sharpened, and hence giving tighter constraints for relaxation rates. 

\section{Sharp bounds for $r$-function}\label{sec:br}

In this section, we prove Theorem \ref{Thm:Main1New} and Theorem \ref{Thm:Main2New}. 
The strategy of the proofs is as follows. We first show the bounds of $r(A,B)$ for the case where $A$ is self-adjoint. 
Then, the decomposition \eqref{eq:CDforA} gives the general bounds for general matrices. (In \ref{sec:App}, we give direct proofs for the case $n=2$.) 
\begin{Proposition}\label{Prop:1}
For any complex matrices $A,B \in M_n(\CA)$ with $A = A^\ct$, 
\begin{equation}\label{eq:kopt2}
\frac{1-\sqrt{2}}{2} \|A\|^2 \|B\|^2 \le r(A,B) \le \frac{1+\sqrt{2}}{2} \|A\|^2 \|B\|^2, 
\end{equation}
where both bounds are sharp. 
\end{Proposition}
[Proof] By the unitary invariance of $r$-function, note that the restriction for $A$ to be self-adjoint is equivalent to the restriction to be real diagonal. Let $A = {\rm diag}[a_1,a_2,\ldots,a_n]$ be any diagonal matrix with real elements $a_i \in \R$ and let $B = (b_{ij})_{i,j=1}^n$ be any complex matrix. 
A direct computation shows
\begin{equation}\label{eq:gDiagA}
r(A,B) = \sum_{i \neq j=1}^n |b_{ji}|^2 (a_i^2 - a_i a_j).
\end{equation}
Notice that the constants $c_\pm := \frac{1 \pm \sqrt{2}}{2}$ appeared in \eqref{eq:kopt2} are solutions of the quadratic equation $4 c (c-1) = 1$. Using this, we have 
\begin{eqnarray*}
&&c_+ \|A \|^2 \|B\|^2 - r(A,B)  \\
&=& c_+ (\sum_k a_k^2) (\sum_{i,j=1}^n |b_{ji}|^2) - \sum_{i \neq j=1}^n |b_{ji}|^2 (a_i - a_i a_j) \\
&\ge& \sum_{i \neq j=1}^n |b_{ji}|^2 ((c_+-1) a_i^2 + c_+ a_j^2 + a_i a_j) \\
&=& \sum_{i \neq j=1}^n |b_{ji}|^2 \Bigl(\sqrt{(c_+-1)}a_i + \sqrt{c_+} a_j\Bigr)^2 \ge 0,
\end{eqnarray*}
where the last completing the square follows from $2 \sqrt{c_+}\sqrt{c_+ -1} = 1$. This shows the upper bound in \eqref{eq:kopt2}. Note that the equality is attained by matrices $A$ and $B$ with e.g., $a_1 = 1 ,a_2 = -\sqrt{\frac{c_+}{(c_+-1)}} = - 2 c_+$ and $b_{12} = 1$ where all other elements are zero. (Here, $B$ should be chosen to be non-normal since for normal $B$ tighter inequality \eqref{eq:NBcase} is satisfied.)  

Similarly, but noting that $c_- < 0$, we have 
\begin{eqnarray*}
&&r(A,B) - c_- \|A \|^2 \|B\|^2   \\
&=& \sum_{i \neq j=1}^n |b_{ji}|^2 (a_i^2 - a_i a_j) + (- c_-) (\sum_k a_k^2) (\sum_{i,j=1}^n |b_{ji}|^2) \\
&\ge& \sum_{i \neq j=1}^n |b_{ji}|^2 ((1 - c_-) a_i^2 + (-c_-) a_j^2 - a_i a_j) \\
&=& \sum_{i \neq j=1}^n |b_{ji}|^2 \Bigl(\sqrt{(1-c_-)}a_i - \sqrt{-c_-} a_j\Bigr)^2 \ge 0,
\end{eqnarray*}
where the last equality follows from $2 \sqrt{-c_-}\sqrt{1- c_-} = 1$. Thus, the lower bound in \eqref{eq:kopt2} is shown. The equality is attained by matrices $A$ and $B$ with e.g., $a_1 = 1 ,a_2 = \sqrt{\frac{-c_-}{(1- c_-)}} = - 2 c_-$ and $b_{12} = 1$ where all other elements are zero.  
\hfill $\blacksquare$

\bigskip

Now we are ready to prove Theorem \ref{Thm:Main1New}. 
\bigskip

\noindent [Proof of Theorem \ref{Thm:Main1New}] Just use decomposition \eqref{eq:CDforA} for the Cartesian decomposition $A = A_R + i A_I$: $r(A,B) = r(A_R,B) + r(A_I,B)$. Noting that $\|A\|^2  = \|A_R\|^2 + \|A_I\|^2$, the application of Proposition \ref{Prop:1} for the self-adjoint $A_R$ and $A_I$ shows that the same bounds \eqref{eq:kopt2} follow for any $A$ and $B$. Moreover, the equalities are attained by choosing $A_R$ and $A_I$ to attain the bounds of \eqref{eq:kopt2}. \hfill $\blacksquare$

\bigskip 

The proof for Theorem \ref{Thm:Main2New} goes similarly. 
\begin{Proposition}\label{Prop:2}
For any complex matrices $A,B \in M_n(\CA)$ with $\tr A = 0$ and $A^\ast = A$,  
\begin{equation}\label{eq:kdiagAna}
\frac{1 - \sqrt{2(1- \frac{1}{n})}}{2}\|A\|^2 \|B\|^2
 \le r(A,B) \le \frac{1 + \sqrt{2(1- \frac{1}{n})}}{2}\|A\|^2 \|B\|^2,
\end{equation}
where both bounds are sharp. 
\end{Proposition}
[Proof] As before, it is enough to prove for any real diagonal matrix $A = {\rm diag}[a_1,a_2,\ldots,a_n]$  ($a_i \in \R$), but this time with the traceless condition $\sum_k a_k = 0$, and for arbitrary $B = (b_{ij})_{i,j=1}^n$. 

The case $n=2$ is straightforward: Letting $A = {\rm diag}[a,-a]$ with $a \in \R$, one finds $
r(A,B) = 2 a^2 (|b_{12}|^2 + |b_{21}|^2)$. This immediately shows the bounds \eqref{eq:kdiagAna} and the bounds are sharp. 
Let $n \ge 3$. Note that constants $c'_\pm := \frac{1 \pm \sqrt{2(1- \frac{1}{n})}}{2}$ appeared in \eqref{eq:kdiagAna} are the solutions of the quadratic equation $4 (c' + \frac{c'}{n-2}-1)(c' + \frac{c'}{n-2}) = (1 + \frac{2c'}{n-2})^2$. 
Using general expression \eqref{eq:gDiagA}, one has
\begin{eqnarray*}
&&c'_+ \|A \|^2 \|B\|^2 - r(A,B) \\
&=& c'_+ (\sum_k a_k^2) (\sum_{i,j=1}^n |b_{ji}|^2) - \sum_{i \neq j=1}^n |b_{ji}|^2 (a_i^2 - a_i a_j) \\
&\ge& \sum_{i \neq j=1}^n |b_{ji}|^2\Bigl( (c'_+-1) a_i^2 + c'_+ a_j^2  +  a_i a_j +  c'_+ \sum_{k \neq i,j} a_k^2 \Bigr). 
\end{eqnarray*}
Applying Schwartz inequality $(n-2) \sum_{k \neq i,j} a_k^2 \ge |\sum_{k \neq i,j} a_k|^2$ and the traceless condition $\sum_{k \neq i,j} a_k = -(a_i + a_j)$,  the last term can be further lower bounded by
\begin{eqnarray*}
&&\sum_{i \neq j=1}^n |b_{ji}|^2\Bigl( (c'_+ + \frac{c'_+}{n-2}-1) a_i^2 + (c'_+ + \frac{c'_+}{n-2}) a_j^2  +  (1 + \frac{2c'_+}{n-2}) a_i a_j\Bigr) \\
&=& \sum_{i \neq j=1}^n |b_{ji}|^2\Bigl(
\sqrt{c'_+ + \frac{c'_+}{n-2}-1} a_i + \sqrt{c'_+ + \frac{c'_+}{n-2}} a_j \Bigr)^2 \ge 0.
\end{eqnarray*}
The last completing the square is due to the fact that $c'_+ = \frac{1 + \sqrt{2(1- \frac{1}{n})}}{2}$ is the solution of $2 \sqrt{c' + \frac{c'}{n-2}-1} \sqrt{c' + \frac{c'}{n-2}} = 1 + \frac{2c'}{n-2}$. 

The equality is also attained by a real diagonal matrix $A$ and $B$ with e.g., 
$a_1 = \sqrt{c'_+ + \frac{c'_+}{n-2}}$, $a_2 = - \sqrt{c'_+ + \frac{c'_+}{n-2}-1}$,
$a_k = \frac{1}{n-2} \left( \sqrt{c'_+ + \frac{c'_+}{n-2}-1} -\sqrt{c'_++ \frac{c'_+}{n-2}}\right)$
 $(3\le k\le n)$ and $b_{21} =1$ where all other elements are zero. 

The lower bound is shown similarly.
Note that $c'_- \le 0$.
Applying 
Schwartz inequality $(n-2) \sum_{k \neq i,j} a_k^2 \ge |\sum_{k \neq i,j} a_k|^2$ and the traceless condition $\sum_{k \neq i,j} a_k = -(a_i + a_j)$, we obtain
\begin{eqnarray*}
&& r(A,B) - c'_- \|A \|^2 \|B\|^2  \\
&=& - c'_- (\sum_k a_k^2) (\sum_{i,j=1}^n |b_{ji}|^2) + 
\sum_{i \neq j=1}^n |b_{ji}|^2 (a_i^2 - a_i a_j) \\
&\ge& \sum_{i \neq j=1}^n |b_{ji}|^2\Bigl( (1- c'_-) a_i^2 - c'_- a_j^2  -  a_i a_j -  c'_- \sum_{k \neq i,j} a_k^2 \Bigr)\\
&\ge&
\sum_{i \neq j=1}^n |b_{ji}|^2\Bigl( (1- c'_- - \frac{c'_-}{n-2}) a_i^2 + (-c'_- - \frac{c'_-}{n-2}) a_j^2  -  (1 + \frac{2c'_-}{n-2}) a_i a_j \Bigr) \\
&=& \sum_{i \neq j=1}^n |b_{ji}|^2\Bigl(
\sqrt{1- c'_- - \frac{c'_-}{n-2}} a_i - \sqrt{-c'_- - \frac{c'_-}{n-2}} a_j \Bigr)^2 \ge 0.
\end{eqnarray*}
The last completing the square follows from $2 \sqrt{1 -c'_- - \frac{c'_-}{n-2}} \sqrt{-c' - \frac{c'_-}{n-2}} = 1 + \frac{2c'_-}{n-2}$.
The equality is attained by 
a real diagonal matrix $A$ and $B$ with e.g., $a_1 = \sqrt{-c'_-- \frac{c'_-}{n-2}}$, $a_2 = \sqrt{1- c'_- - \frac{c'_-}{n-2}}$,
$a_k = \frac{1}{n-2} \left(- \sqrt{1-c'_- - \frac{c'_-}{n-2}} -\sqrt{-c'_-- \frac{c'_-}{n-2}}\right)$
 $(3\le k\le n)$ and $b_{21}=1$ where all other elements are zero. \hfill $\blacksquare$

\bigskip 

\noindent [Proof of Theorem \ref{Thm:Main2New}] 
Similar to the proof of Theorem \ref{Thm:Main1New}, this follows from Proposition \ref{Prop:2} and the decomposition \eqref{eq:CDforA} just by noting that  both the real and imaginary parts are also traceless for a traceless matrix $A$. The sharpness of the bounds also follows similarly. 
\hfill $\blacksquare$ 

\section{Application to quantum dynamical semigroup}\label{sec:Phys}

Note that the trivial condition that any relaxation rate \eqref{eq:DefRC} is positive can be guaranteed by the positive preserving property of the dynamical map. However, it is known that the condition of complete positivity imposes a strong constraint on the relaxation rates \cite{ref:GKS,ref:GK,ref:KAW,ref:CKKS}: 
Simply put, any relaxation rate cannot be too large compared to other relaxation rates. This is quantitatively described by the following constraint: 	
\begin{equation}\label{eq:Restriction}
\Gamma_\alpha \le r(n) \sum_{\beta = 1}^{n^2-1} \Gamma_\beta \quad (\forall \alpha = 1,\ldots,n^2-1)
\end{equation}
with some positive constant $r(n)$ dependent on the level $n$. Note that the condition $r(n) < 1$ yields a non-trivial constraint for relaxation rates and the smaller the constant is the tighter the constraint becomes. Therefore, it is an important problem to find the minimum constant $r_{\rm opt}(n)$ where the constraint \eqref{eq:Restriction} holds for any GKLS generator.
In light of the universality of the property, the constraint will give a physical manifestation of a mathematical condition of complete positivity \cite{ref:CKKS}.

For $n=2$ (i.e., qubit system), we have shown \cite{ref:GKS,ref:GK} that for any GKLS generator $r(2) = \frac{1}{2}$ is correct for the bound \eqref{eq:Restriction}.  The constraint can be rephrased as
$$
\Gamma_i \le \Gamma_j + \Gamma_k \quad (i,j,k = 1,2,3).
$$
Interestingly, in the case where two of the relaxation rates are equal, say $\Gamma_2 = \Gamma_3$, the relation coincides with the famous relation between the longitudinal relaxation time $T_L (=1/\Gamma_1)$ and the transverse relaxation time $T_T (=1/\Gamma_2 = \Gamma_3)$:
\begin{equation}\label{eq:TlTT}
  T_{L} \geq 2 T_T,
\end{equation}
which is experimentally demonstrated to be true \cite{ref:Alicki,ref:LT}.
For a general $n$, not much is known.
Applying the result from \cite{ref:WC}, one finds $r(n) \le \frac{2}{n}$ but only for the restricted class of purely dissipative generator. In \cite{ref:KAW}, we have shown that $r(n) = \frac{\sqrt{2}}{n}$ is correct for any GKLS generator with general $n$ by using BW-inequality. On the other hand, we have shown that the best constant is lower bounded by $\frac{1}{n}$ \cite{ref:CKKS}. Combining these results, we can conclude
\begin{equation}\label{eq:qubitRC}
r_{\rm opt}(2) = \frac{1}{2}.
\end{equation}
But, for larger $n \ge 3$, we only know
\begin{equation}\label{eq:nRC}
\frac{1}{n} \le r_{\rm opt}(n) \le \frac{\sqrt{2}}{n}.
\end{equation}

We show here that the bound problem \eqref{eq:bd} of $r$-function can be used to tackle on this problem. 
Let us first show the general expression \eqref{eq:RRRF} for relaxation rate. 
Applying GKLS form \eqref{DIAG} into the eigenvalue equation $\LA(u_\alpha) =\lambda_\alpha u_\alpha$ with unit eigenmatrix $u_\alpha$, and multiplying $u^\ct_\alpha$ to the equation from the left, and finally taking the trace, we obtain
$$
\lambda_\alpha = \tr \Bigl( u^\ct_\alpha(-i[H,u_\alpha] + \frac{1}{2}
\sum_{k}(2 L_k u_\alpha L_k^\ct - L_k^\ct L_k u_\alpha - u_\alpha L_k^\ct L_k)) \Bigr).
$$
Taking the real part of this equation and noting $\Gamma_\alpha = - \Re \lambda_\alpha$, one arrives at the expression \eqref{eq:RRRF} especially if one uses the form \eqref{eq:r-expand}. 
On the other hand, we know the relation (see \cite{ref:KAW,ref:CKKS}):
$$
\sum_{\alpha = 1}^{n^2-1} \Gamma_\alpha= n \sum_k \| L_k \|^2.
$$
In showing this, we need to use the traceless condition for $L_k$. 
However, this can be assumed without loss of generality as the trace part of $L_k$ can be renormalized into the Hamiltonian part in the generator \eqref{DIAG}.  
Combining these results, the bound of the form \eqref{eq:bd} gives a non trivial constraint \eqref{eq:ConsRelCN} for relaxation rate. Moreover, the trace of $u_\alpha$ for non-zero eigenvalue $\lambda_\alpha$ is traceless. This is easily shown by using the fact ${\rm tr}(\LA(A)) = 0$ 
for any matrix $A$.
Therefore, the bound \eqref{eq:bd} with the restriction of $A$ being traceless also gives a non-trivial constraint for the relaxation rates. 
Finally, by applying results of Theorem \ref{Thm:Main1New} and \ref{Thm:Main2New}, we obtain 
\begin{Theorem}\label{Thm:MAIN}
For $n$-level quantum system, the optimal bound for relaxation rates for any quantum dynamical semigroup satisfies
\begin{equation}
r_{\rm opt}(n) \le \frac{1 + \sqrt{2(1- \frac{1}{n})}}{2n} \le \frac{1 + \sqrt{2}}{2n}.
\end{equation}
\end{Theorem}
We emphasize that the right inequality due to Theorem \ref{Thm:Main1New} already gives tighter constraint than previously known bound \eqref{eq:nRC}. 

\section{Conclusion and discussion}\label{sec:CD}

In this paper, we have introduced a $r$-function \eqref{eq:rf} with which any relaxation rate of quantum dynamical semigroup is expressed by \eqref{eq:RRRF}. 
We discussed the upper and lower bounds of form \eqref{eq:bd} and found the best bounds are given by $c_\pm = \frac{1\pm\sqrt{2}}{2}$ independent of the matrix size (Theorem \ref{Thm:Main1New}). If we restrict the matrix $A$ to be traceless, the bounds are reduced to be $c_\pm = \frac{1\pm\sqrt{2(1-\frac{1}{n})}}{2}$ (Theorem \ref{Thm:Main2New}).  
As an application, we obtained tighter universal constraints on relaxation rates than any of previously known constraints (Theorem \ref{Thm:MAIN}). 

In \cite{ref:CKKS}, we raised a conjecture that $r_{\rm opt}(n) = \frac{1}{n}$ for any $n \ge 2$. 
The present result, especially the upper bound $\frac{1 + \sqrt{2(1- \frac{1}{n})}}{2n}$ is close to this and indeed attained for $2$-level (i.e., qubit) system. Unfortunately, for $n \ge 3$, this is strictly larger than $\frac{1}{n}$, hence still leaves this problem open.

\bigskip

\noindent {\bf Acknowledgements}

\bigskip 

We would like to thank Prof. Andrzej Kossakowski for his fruitful comments and advices.
During the preparation of this manuscript, he passed away on 1th FEB 2021.
We would like to dedicate this paper to his memory. D.C.  is supported by the National Science Center project 2018/30/A/ST2/00837.
G. K. is supported in part by JSPS KAKENHI Grants No. 17K18107.

\appendix

\section{Direct proofs of Theorems \ref{Thm:Main1New} and \ref{Thm:Main2New} for $n=2$. }\label{sec:App}

In this appendix, we present direct proofs for Theorem \ref{Thm:Main1New} and Theorem \ref{Thm:Main2New} for the case $n=2$ only for instructive purpose. Namely, for any complex matrices $A,B \in M_2(\CA)$, 
\begin{equation}\label{eq:kopt2-2}
\frac{1-\sqrt{2}}{2} \|A\|^2 \|B\|^2 \le r(A,B) \le \frac{1+\sqrt{2}}{2} \|A\|^2 \|B\|^2.  
\end{equation}
If $A$ is restricted to be traceless, 
\begin{equation}\label{eq:topt2}
0 \le r(A,B) \le \|A\|^2 \|B\|^2.  
\end{equation}
In both cases, the bounds are sharp. 

In the following, we use standard notations (in vector analysis) for complex vectors ${\bm a},{\bm b} \in \CA^3$ such as ${\bm a} \cdot {\bm b} := \sum_{i=1}^3 \ov{a_i} b_i$ (dot product), $|{\bm a}| = \sqrt{{\bm a} \cdot {\bm a}} = \sum_{i=1}^3 |a_i|^2$, and ${\bm a} \times {\bm b}$ (cross product) which is defined by $({\bm a} \times {\bm b})_i = \sum_{j,k=1}^3 \epsilon_{ijk} a_j b_k  \ (i=1,2,3)$ where $\epsilon_{ijk}$ is the Levi-Civita symbol. 
We start from the following elementary lemmas: As in the main text, let $c_\pm := \frac{1 \pm \sqrt{2}}{2}$ which are the solutions of $4c(c-1) = 1$. 
\begin{Lemma}\label{Lem:k2}
Let ${\bm a} = (a_1,a_2,a_3)$ and ${\bm b}=(b_1,b_2,b_3)$ be complex vectors in $\CA^3$ with ${\bm b} = {\bm b}_R + i {\bm b}_I$ (the real and the imaginary parts). Then, for any $x \in \R$, we have 
\begin{equation}\label{eq:cg2comp}
c_+ (x^2 + |{\bm a}|^2)|{\bm b}|^2  \ge |{\bm a}|^2 |{\bm b}|^2 + 2 x |{\bm a}||{\bm b}_R| |{\bm b}_I|.
\end{equation}
In case ${\bm b} \neq 0$, the equality is attained at $x = (\sqrt{2}-1)|{\bm a}|$ if $|{\bm b}_R|=|{\bm b}_I|$.
\end{Lemma}
[Proof] If ${\bm b} = 0$, the inequality trivially holds. Assuming ${\bm b} \neq 0$, inequality \eqref{eq:cg2comp} is simply the following quadratic inequality for $x$: 
$$
c_+ |{\bm b}|^2 x^2 - 2 |{\bm a}||{\bm b}_R| |{\bm b}_I| x + (c_+-1) |{\bm a}|^2|{\bm b}|^2 \ge 0, 
$$
hence is equivalent to the negative semi-definiteness of its discriminant:
$$
0 \ge  (2 |{\bm a}||{\bm b}_R| |{\bm b}_I|)^2 - |{\bm b}|^2 |{\bm a}|^2|{\bm b}|^2, 
$$
where we used the fact that $c_+$ is one of the solutions of $4c(c-1) = 1$. 
Since $| {\bm b}|^2 = |{\bm b}_R|^2 + |{\bm b}_I|^2$, the right hand side is $- |{\bm a}|^2 (|{\bm b_R}|^2 - |{\bm b_I}|^2)^2 \le 0$. This proves the inequality \eqref{eq:cg2comp} for all $x$. Finally, the discriminant is zero when $|{\bm b_R}| = |{\bm b_I}|$. Thus, the equality for \eqref{eq:cg2comp} is satisfied at the point $x = \frac{2 |{\bm a}||{\bm b}_R| |{\bm b}_I|}{2 c_+ |{\bm b}|^2} = \frac{|{\bm a}|}{2 c_+} = (\sqrt{2}-1) |{\bm a}|$. \hfill $\blacksquare$

\bigskip

\begin{Lemma}\label{Lem:k2Lower}
For any $x \in \R$, ${\bm y},{\bm z}, {\bm w} \in \CA^3$, we have 
\begin{equation}\label{eq:cg2comp2}
(|{\bm y}|^2+|{\bm z}|^2)(c_+ |{\bm w}|^2 - c_- x^2) \ge 2 |{\bm w}| |{\bm y}| |{\bm z}| x. 
\end{equation}
In case $|{\bm y}|^2+|{\bm z}|^2 \neq 0$, the equality is attained if $x = (\sqrt{2}+1)|{\bm w}| $ if $|{\bm y}|^2 =|{\bm z}|$. 
\end{Lemma}
[Proof] If ${\bm y} = {\bm z} = 0$, inequality \eqref{eq:cg2comp2} trivially holds. Assuming $|{\bm y}|^2+|{\bm z}|^2 \neq 0$, the inequality is the following quadratic inequality: 
$$
-c_- (|{\bm y}|^2+|{\bm z}|^2)x^2 - 2 |{\bm w}| |{\bm y}| |{\bm z}| x + c_+ (|{\bm y}|^2+|{\bm z}|^2) |{\bm w}|^2 \ge 0,  
$$
which is equivalent to the non-positivity of the discriminant:
$$
4|{\bm w}|^2 |{\bm y}|^2 |{\bm z}|^2 + 4 c_+ c_-  (|{\bm y}|^2+|{\bm z}|^2)^2 |{\bm w}|^2 \le 0. 
$$
However, noting $c_+ c_- = - \frac{1}{4}$, the left hand side is $- |{\bm w}|^2(|{\bm y}|^2 - |{\bm z}|^2)^2$, hence the non-positivity is satisfied. In particular, the discriminant is zero if $|{\bm y}| = |{\bm z}|$ and the equality is attained at $x = - \frac{|{\bm w}|}{2c_-} = (\sqrt{2}+1) |{\bm w}|$. 
\hfill $\blacksquare$ 

\bigskip 

[Proofs of \eqref{eq:kopt2-2} and \eqref{eq:topt2}] We use the orthonormal basis $F_0 = \I/\sqrt{2}, F_i = \sigma_i/\sqrt{2} \ (i=1,2,3)$ of $M_2(\CA)$ where $\sigma_1,\sigma_2,\sigma_3$ are Pauli matrices:
$$
\sigma_1=\left(
\begin{array}{cc}
 0 & 1 \\
 1 & 0 \\
\end{array}
\right),
\sigma_2=\left(
\begin{array}{cc}
 0 & -i \\
 i & 0 \\
\end{array}
\right), \sigma_3
=\left(
\begin{array}{cc}
 1 & 0 \\
 0 & -1 \\
\end{array}
\right).
$$
Arbitrary $2\times 2$ matrices $A$ and $B$ are written by $A =  \sum_{\mu=0}^3 a_\mu F_\mu $ and $B = \sum_{\mu=0}^3 b_\mu F_\mu$ with $a_\mu,b_\mu \in \CA$.
Note that, for the case of \eqref{eq:topt2}, one simply uses $a_0 = 0$ for traceless condition for $A$.
One has $\| A\|^2 = |a_0|^2 + |{\bm a}|^2, \| B\|^2 = |b_0|^2 + |{\bm b}|^2$.
A direct computation shows
\begin{equation}\label{eq:gBex}
r(A,B) = |{\bm a}|^2 |{\bm b}|^2 - \frac{1}{2}(|{\bm a} \cdot {\bm b}|^2 + |\ov{{\bm a}} \cdot {\bm b} |^2) -  {\rm Im} (\ov{a_0} {\bm a} \cdot (\ov{\bm b} \times {\bm b}) ),
\end{equation}
where ${\bm a} = (a_1,a_2,a_3)$ and ${\bm b}=(b_1,b_2,b_3)$ are complex three dimensional vectors.
Therefore, if we restrict $A$ to be traceless, i.e., $a_0 = 0$, the third terms in \eqref{eq:gBex} vanishes:
$$
r(A,B) = |{\bm a}|^2 |{\bm b}|^2 - \frac{1}{2}(|{\bm a} \cdot {\bm b}|^2 + |\ov{{\bm a}} \cdot {\bm b} |^2). 
$$
By Scwarz inequality, this is lower bounded by $0$. Also, one has 
\begin{equation}\label{eq:gBexTr0}
r(A,B)  \le |{\bm a}|^2 |{\bm b}|^2 \le (|a_0|^2 + |{\bm a}|^2) (|b_0|^2 + |{\bm b}|^2) = \|A\|^2 \|B\|^2.  
\end{equation}
Moreover, it is easy to construct matrices $A$ and $B$ to attain all the above equalities: 
For the lower bound, take parallel real vectors ${\bm a}$ and ${\bm b}$.  For the upper bound, generally, any $A$ and $B$ with $a_0=b_0=0$ and orthogonal conditions ${\bm a} \cdot {\bm b} = \ov{{\bm a}} \cdot {\bm b} = 0$ attain the bound. 
In particular, if we use real ${\bm a}$, the equality is achieved by a self-adjoint matrix $A$. This completes the proof of \eqref{eq:topt2}. 

To show the general bounds \eqref{eq:kopt2-2}, let ${\bm b} = {\bm b}_R + i {\bm b}_I$, so that $\ov{\bm b} \times {\bm b} = 2 i {\bm b}_R \times {\bm b}_I$, and therefore  
\begin{equation}\label{eq:gBex2}
r(A,B) = |{\bm a}|^2 |{\bm b}|^2 - \frac{1}{2}(|{\bm a} \cdot {\bm b}|^2 + |\ov{{\bm a}} \cdot {\bm b} |^2) - 2 {\rm Re} (\ov{a_0} {\bm a} \cdot ({\bm b}_R \times {\bm b}_I) ).
\end{equation}
Using this expression, one observes $
\frac{1 + \sqrt{2}}{2} \|A\|^2 \|B\|^2 - r(A,B) \ge \frac{1 + \sqrt{2}}{2} (|a_0|^2 + |{\bm a}|^2)|{\bm b}|^2 - |{\bm a}|^2 |{\bm b}|^2 - 2 |a_0| | {\bm a}| |{\bm b}_R| | {\bm b}_I|$ where use has been made of Schwarz inequality and $|{\bm b}_R \times {\bm b}_I| \le |{\bm b}_R||{\bm b}_I|$. However, by Lemma \ref{Lem:k2} (for $x = |a_0|$), this is non-negative, hence, one has the upper bound in \eqref{eq:kopt2-2}. 
It is easy to construct matrices $A$ and $B$ to achieve the equality. 
For instance, let $\{{\bm a},{\bm b_R},{\bm b_I}\}$ form a left-handed orthonormal base of $\R^3$. Then, using the equality condition in Lemma \ref{Lem:k2}, all equalities in the above inequalities are attained by taking $a_0 = \sqrt{2}-1$ and $b_0 = 0$. 

Finally, to show the lower bound in \eqref{eq:kopt2-2}, let $V = {\rm span}_{\mathbb C} \{ {\bm b}_R , {\bm b}_I \}$ and $P$ be a projection onto $V$,
and let $P^\perp = I - P$.
Then, $P{\bm b} = {\bm b}$, $P\overline{\bm b} = \overline{\bm b}$ 
and $P^\perp ({\bm b}_R \times {\bm b}_I) ={\bm b}_R \times {\bm b}_I$.
Therefore,
$
|{\bm a} \cdot {\bm b}| = |{\bm a} \cdot P{\bm b}| =|P{\bm a} \cdot {\bm b} | 
\quad {\rm and} \quad
|\overline{\bm a} \cdot {\bm b}| =|{\bm a} \cdot \overline{\bm b}| =
| P{\bm a} \cdot \overline{\bm b} |
$, and $
{\bm a} \cdot ({\bm b}_R \times {\bm b}_I )  =
 P^\perp {\bm a} \cdot ({\bm b}_R \times {\bm b}_I)$. 
Considering $|{\bm a}|^2 = |P{\bm a}|^2 + |P^\perp {\bm a}|^2$, we have $r(A,B) - \frac{1-\sqrt{2}}{2} \|A\|^2\|B\|^2 \ge \frac{\sqrt{2}-1}{2}|{a}_0|^2 |{\bm b}|^2 
+\frac{\sqrt{2}+1}{2}|P^\perp {\bm a}|^2 |{\bm b}|^2 
-2 {\rm Re} (\overline{a_0} (P^\perp{\bm a})\cdot ({\bm b}_R \times {\bm b}_I)) \ge  \frac{\sqrt{2}-1}{2}|{a}_0|^2 |{\bm b}|^2 +\frac{\sqrt{2}+1}{2}|P^\perp {\bm a}|^2 |{\bm b}|^2 
-2 |a_0| |P^\perp{\bm a}||{\bm b}_R||{\bm b}_I|$. 
This is shown to be non-negative by using Lemma \ref{Lem:k2Lower} for $x = |a_0|, {\bm y} = {\bm b}_R, {\bm z} = {\bm b}_I, {\bm w} = P^\perp {\bm a}$, hence we have the lower bound in \eqref{eq:kopt2-2}. 
It is also easy to construct matrices $A$ and $B$ to achieve the equality. 
For instance, again let $\{{\bm a},{\bm b_R},{\bm b_I}\}$ form a left-handed orthonormal base of $\R^3$. 
Then, using the equality condition in Lemma \ref{Lem:k2Lower}, all equalities in the above inequalities are attained by taking $a_0 = \sqrt{2}+1$ and $b_0 = 0$. 


\hfill $\blacksquare$

\bigskip


\begin{thebibliography}{1} \bibliographystyle{plain}


	

\bibitem{ref:BW1} A. B\"{o}ttcher, D. Wenzel, The Frobenius norm and the commutator, Linear Algebra Appl. {\bf 429}, 1864 (2008).

\bibitem{ref:Breuer} H.~P. Breuer and F. Petruccione, {\em The Theory of Open Quantum Systems} (Oxford Univ. Press, Oxford, 2007).

\bibitem{ref:Dav} E.~B. Davies, {\em Quantum Theory of Open Systems} (Academic Press, London, 1976).

\bibitem{ref:Alicki} R. Alicki and K. Lendi, {\it Quantum Dynamical Semigroups and Applications} (Springer, Berlin, 1987).


\bibitem{ref:GZ} C.~W. Gardiner and P. Zoller, {\em Quantum Noise}, 2nd ed. (Springer-Verlag, Heidelberg, 2000). 

\bibitem{ref:GKreview} G. Kimura, {\it Elementary mathematical framework for open quantum $d$-level systems: Decoherence Overview}, in: M. Nakahara, R. Rahimi, A. SaiToh (eds.), {\it Decoherence Suppression in Quantum Systems 2008}, World Scientific Publishers, 2009, 1. 

\bibitem{ref:GKS} V. Gorini, A. Kossakowski, and E. C. G. Sudarshan, J. Math. Phys. {\bf 17}, 821 (1976). 
\bibitem{ref:L}  G. Lindblad, Comm. Math. Phys. {\bf 48}, 119 (1976).

\bibitem{ref:GK} G. Kimura, Phys. Rev. A {\bf 66}, 062113 (2002).

\bibitem{ref:KAW} G. Kimura, S. Ajisaka, K. Watanabe, Universal Constraints on Relaxation Times for $d$-Level GKLS Master Equations, Open Syst. Inform. Dynam. 24, 1 (2017).

\bibitem{ref:CKKS} D. Chruscinski, G. Kimura, A. Kossakowski, Y. Shishido, On the universal constraints for relaxation rates for quantum dynamical semigroup, arXiv:2011.10159. 

\bibitem{ref:Dirac} P.~A.~M. Dirac, The Principles of Quantum Mechanics (Clarendon Press; 1982). 

\bibitem{ref:BW0} A. B\"{o}ttcher, D. Wenzel, How big can the commutator of two matrices be and how big is it typically? Linear Algebra Appl. 403, 216 (2005). 

\bibitem{ref:BW2} S. W. Vong, X. Q. Jin, Proof of B\"{o}ttcher and Wenzel's conjecture, Oper. Matrices, {\bf 2},  435 (2008).

\bibitem{ref: Lu} Z. Lu, Normal Scalar Curvature Conjecture and its applications, J. Funct. Anal., {\bf 261}, 1284 (2007).

\bibitem{ref:BW_Var} K. M. R. Audenaert, Variance bounds, with an application to norm bounds for commutators, Linear Algebra Appl. {\bf 432}, 1126 (2009).


\bibitem{ref:Comm_maxFrob} C. M. Cheng, S. W. Vong, D. Wenzel, Commutators with maximal Frobenius norm, Linear Algebra Appl. {\bf 432}, 292 (2010).

\bibitem{ref:Bhatia} R. Bhatia, {\em Matrix Analysis} (Springer 1997).  

\bibitem{ref:Kraus} K. Kraus, {\em States, Effects, and Operations} (Springer, 1983). 

\bibitem{ref:NC} M.~A. Nielsen, I. L.~Chuang, {\em Quantum Computation and Quantum Information} (Cambridge, England, 2000).

\bibitem{ref:GKTextbook} M. Hayashi, S. Ishizaka, A. Kawachi, G. Kimura, T. Ogawa, {\em Introduction to Quantum Information Science} (Springer, 2015).


\bibitem{ref:WC} M.~M. Wolf, J.~I. Cirac, Commun. Math. Phys. {\bf 279}, 147 (2008). 

\bibitem{ref:LT} A. Abragam, \textit{ Principles of Nuclear Magnetism} (Oxford University Press, 1961);  C. P. Slichter, \textit{ Principles of Magnetic Resonance} (Springer-Verlag, 1990).
\end{thebibliography}
\end{document}